\documentclass[reprint,superscriptaddress, amsmath,amssymb, aps,prb,floatfix]{revtex4-2}

\usepackage{graphicx}
\usepackage{bm}
\usepackage[version=4]{mhchem}
\usepackage{xcolor}

\begin{document}


\title{Raman magnon spectroscopy of local interactions and ground state selection in \ce{Sr2IrO4}}

\author{Xiang Li}
\affiliation{Department of Physics, University of Toronto, Toronto, Ontario M5S 1A7, Canada}
\affiliation{Department of Chemistry, University of Toronto, Toronto, Ontario M5S 1A7, Canada}
\affiliation{Division of Physics, Math, and Astronomy, California Institute of Technology, Pasadena, California 9115, USA}
\affiliation{Okinawa Institute of Science and Technology Graduate University, Onna, Okinawa 904-0495, Japan}
\email{Corresponding authors: xianglll.li@utoronto.ca, dmsilev@caltech.edu}

\author{Scott E. Cooper}
\affiliation{Okinawa Institute of Science and Technology Graduate University, Onna, Okinawa 904-0495, Japan}

\author{Ahmed E. Fahmy}
\affiliation{Department of Physics, The Ohio State University, Columbus, Ohio 43210, USA}

\author{Yuan-Ming Lu}
\affiliation{Department of Physics, The Ohio State University, Columbus, Ohio 43210, USA}

\author{A. de la Torre}
\affiliation{Division of Physics, Math, and Astronomy, California Institute of Technology, Pasadena, California 9115, USA}
\affiliation{Department of Physics, Northeastern University, Boston, Massachusetts 02115, USA}
\affiliation{Quantum Materials and Sensing Institute, Northeastern University Innovation Campus, Burlington, MA 01803, USA.}

\author{R. S. Perry}
\affiliation{London Centre for Nanotechnology and Department of Physics and Astronomy, University College London, London WC1E 6BT, United Kingdom  }
\affiliation{ISIS Facility, Rutherford Appleton Laboratory, Didcot OX11 0QX, United Kingdom }

\author{D. Hsieh}
\affiliation{Division of Physics, Math, and Astronomy, California Institute of Technology, Pasadena, California 9115, USA}

\author{T.F. Rosenbaum}
\affiliation{Division of Physics, Math, and Astronomy, California Institute of Technology, Pasadena, California 9115, USA}

\author{Yejun Feng}
\affiliation{Division of Physics, Math, and Astronomy, California Institute of Technology, Pasadena, California 9115, USA}
\affiliation{Okinawa Institute of Science and Technology Graduate University, Onna, Okinawa 904-0495, Japan}

\author{D.M. Silevitch}
\affiliation{Division of Physics, Math, and Astronomy, California Institute of Technology, Pasadena, California 9115, USA}

\date{\today}

\begin{abstract}
Competing and coupled spin and charge interactions in quantum materials lead to a variety of ordered states where local configurations preferentially influence the long-range order. When these interactions are finely balanced in energy, disorder and fluctuations play an outsized role. Raman scattering is particularly well-suited to revealing the underlying physics in such situations because of its sensitivity to local environments and ability to reveal overall symmetries. We perform angle-resolved Raman polarization measurements on single crystals of the correlated, layered magnet, \ce{Sr2IrO4}, where the Mott insulating ground state arises from strong spin-orbit coupling. We characterize the symmetries of both the phonon and magnon modes through comprehensive measurements in both the $ab$-plane and out-of-plane geometries from 10 to 700 cm$^{-1}$, and trace the evolution of these modes in both configurations to 12 GPa in a diamond anvil cell with perforated diamonds. Pressure does not significantly alter the lattice as the phonon modes shift linearly under compression, but at the same time the magnon modes become position dependent and spread over a range of wavenumbers. We attribute this magnetic heterogeneity to pressure-enhanced variations in the weak interlayer interactions, which may locally favor competing magnetic stacking configurations, and compare our experimental results to the predictions of linear spin wave calculations. Our results demonstrate that Raman-active magnons amplify $\mu$eV-scale interactions responsible for ground-state selection into easily measurable spectral changes.   
\end{abstract}

\maketitle


\section{Introduction}

In the presence of finely balanced interactions and competing ground states, global long-range order can harbor a variety of local configurations. Disorder can amplify these effects, especially in the quantum regime  \cite{savaryDisorderInducedQuantumSpin2017}. Experimental manifestations in magnetic systems include Dzyaloshinskii-Moriya (DM) induced non-collinear spin arrangements, power-law decay of short-range order, and dynamical and coherent states of quantum spin liquids. Typically, the degeneracy of magnetic ground states is induced by frustration, with an overlay of disorder. 

One particularly rich realization involves two-dimensional (2D) sheets of magnetic moments with weak interplane interactions controlling the stacking along the out-of-plane direction \cite{Cao:2016ep}. To that end, we explore lattice and magnetic excitations in the layered antiferromagnetic Mott insulator \ce{Sr2IrO4}, with an emphasis on general symmetry considerations and the plethora of possible local spin stacking configurations. Results under hydrostatic pressure along different crystalline axes reveal the evolving influence of disorder.

Disorder can exacerbate the experimental challenge with stacking faults determining different ground states (viz. \ce{\alpha-RuCl3} \cite{Cao:2016ep}). Weak interactions can play an outsized role when couplings compete, a scenario that becomes especially pronounced at quantum phase transitions \cite{laughlinQuantumCriticalityConundrum2001} and in quantum spin liquids. Moreover, the large degeneracy in stacking configurations of comparable energy can be lifted through a potential order-from-disorder mechanism driven by quantum fluctuations, even when the further neighbor interactions approach zero \cite{chubukovOrderDisorderKagome1992}. Differentiating between a variety of exotic mechanisms thus requires a detailed characterization and analysis of the physical system under investigation.  

Originally investigated as a relative to the cuprate superconductors, it was soon realized that \ce{Sr2IrO4}’s Mott insulator ground state derives from strong spin-orbit coupling \cite{kimNovel$J_mathrmeff12$2008}. The antiferromagnetism arises from the organization of Ir $5d$ states into effective $J_\mathrm{eff} = 1/2$ and $J_\mathrm{eff} = 3/2$ manifolds with pronounced electron–electron interactions  \cite{kimNovel$J_mathrmeff12$2008,kimPhaseSensitiveObservationSpinOrbital2009}. Crucially, staggered rotations of the \ce{IrO6} octahedra  lower the crystal symmetry and generate DM interactions, producing canted antiferromagnetism with a weak ferromagnetic moment in each \ce{IrO2} layer \cite{crawfordStructuralMagneticStudies1994}. The stacking of these weak moments along the crystallographic $c$-axis renders interlayer coupling an essential ingredient of the magnetic ground state, despite the quasi-2D electronic structure. Stacking configurations and layer number also have been recognized as key factors in determining the magnetic ground state in examples such as \ce{CrI3} \cite{songSwitching2DMagnetic2019,liPressurecontrolledInterlayerMagnetism2019} and \ce{MnBi2Te4} \cite{padmanabhanInterlayerMagnetophononicCoupling2022,klimovskikhTunable3D2D2020}.

This coupling between structure and magnetism makes \ce{Sr2IrO4} unusually responsive to external perturbations. Modest changes in lattice geometry or symmetry can induce pronounced changes in magnetic order and low-energy excitations, reflecting the delicate balance among exchange interactions, spin–orbit coupling, and interlayer stacking. As a result, pressure, epitaxial strain, dimensional confinement, and nonequilibrium excitation have all been shown to strongly modify the insulating and magnetic states, underscoring the close energetic competition among distinct phases in \ce{Sr2IrO4} \cite{parisStrainEngineeringCharge2020,haskelPossibleQuantumParamagnetism2020,choiLightinducedInsulatorMetal2024}. Under pressure, it is generally believed that the antiferromagnetism is suppressed at 20 GPa  \cite{haskelPossibleQuantumParamagnetism2020,liMagneticOrderDisorder2021}, while the system remains insulating to at least 185 GPa  \cite{Zocco:2014bd,chenPersistentInsulatingState2020}, opening the possibility of a potential spin liquid state hosted by a Mott insulator.

The ambient-pressure lattice of \ce{Sr2IrO4} is generally recognized to be tetragonal with $a=b\sim5.5$~Å and $c\sim 26 $~Å, but the space group is less clear. Initial studies suggested that the lattice had a space group of $I4_1/acd$ with point group $D_{4h}$ ($4/mmm$) \cite{crawfordStructuralMagneticStudies1994}. However, second-harmonic generation (SHG) \cite{torchinskyStructuralDistortionInducedMagnetoelastic2015} and more recent neutron diffraction observations of forbidden reflections of $(m, 0, l)$ type with $m$ and $l$ both odd \cite{yeMagneticCrystalStructures2013,yeStructureSymmetryDetermination2015} suggest a space group of $I4_1/a$ with a point group of $C_{4h}$ ($4/m$) \cite{torchinskyStructuralDistortionInducedMagnetoelastic2015,yeStructureSymmetryDetermination2015}. With eight chemical units in the primitive cell, \ce{Sr2IrO4} in the $D_{4h}$ point group would allow 13 Raman-active modes (3 $A_{1g}$ + 5 $B_{1g}$ + 4 $B_{2g}$ + 1 $E_g$) \cite{gretarssonRamanScatteringStudy2017}, but several accounts in the literature have reported observations of four $A_{1g}$ modes between 100 and 600 cm$^{-1}$ \cite{liMagneticOrderDisorder2021,gretarssonRamanScatteringStudy2017}. 

While the $I4_1/a$ lattice space group necessarily introduces two inequivalent Ir sites \cite{yeStructureSymmetryDetermination2015}, the commonly accepted magnetic structure \cite{yeMagneticCrystalStructures2013} employed the point group $C_{2h}$ ($mmm$) which has no group-subgroup relationship to $C_{4h}$ and assumes one Ir site. This simplification is mostly due to the limited number of available neutron magnetic reflections for the refinement \cite{YePrivate2026}. The Ir spins generally order within the $ab$ plane, breaking the four-fold in-plane rotational symmetry  \cite{yeMagneticCrystalStructures2013,boseggiaLockingIridiumMagnetic2013}. DM interactions generate a weak ferromagnetic moment in each \ce{IrO2} layer \cite{crawfordStructuralMagneticStudies1994,jackeliMottInsulatorsStrong2009}. There exist many proposed schemes for the sheet stacking along the out-of-plane $c$-axis \cite{porrasPseudospinlatticeCouplingSpinorbit2019}, with nearly degenerate energies \cite{carterTheoryMetalinsulatorTransition2013}. This degeneracy reflects strong symmetry-imposed frustration of interlayer couplings, which suppresses the leading isotropic and DM exchange terms between adjacent layers. Hence magnetic ground-state selection in \ce{Sr2IrO4} is highly sensitive to subleading anisotropic interactions and longer-range interlayer terms. Such interactions lie at energy scales far smaller than the dominant \emph{in-plane} exchange interaction \cite{takayamaModelAnalysisMagnetic2016}. These considerations underscore the importance of experimental probes that can directly access the role of interlayer coupling in magnetic ground-state selection.

As excitations provide a high-fidelity reflection of the ground state, magnon modes are potentially sensitive to the sheet stacking and other long-range magnetic ordered structures. Historically, optical Raman scattering has been used widely to explore magnon modes, with application to \ce{Sr2IrO4}  \cite{liMagneticOrderDisorder2021,gretarssonRamanScatteringStudy2017,gimIsotropicAnisotropicRegimes2016,liOpticalRamanMeasurements2020}. For high-quality samples of \ce{Sr2IrO4} at ambient pressure, there appears to be only one magnon Raman mode at $\sim18-19$ cm$^{-1}$ \cite{liMagneticOrderDisorder2021,gretarssonRamanScatteringStudy2017,gimIsotropicAnisotropicRegimes2016}. This mode was observed in spectra measured with light propagating parallel to the surface normals of both the $ab$-plane and the side-plane and was attributed to $B_{2g}$ in the $D_{4h}$ setting \cite{gretarssonRamanScatteringStudy2017}. An extra peak at $\sim10$ cm$^{-1}$ has been reported occasionally in \ce{Sr2IrO4} under ambient conditions \cite{gimIsotropicAnisotropicRegimes2016}, but it is sample dependent, suggestive of the influence of disorder introduced by either non-stoichiometry \cite{kimSingleCrystalGrowth2022} or residual strain. 

A multitude of magnon modes have been observed under pressure in a diamond anvil cell \cite{liMagneticOrderDisorder2021}, and they are found to be dependent on sample position and random in energy. Surprisingly, magnon modes measured from the side plane all demonstrate blue shifts relative to the ambient pressure 18 cm$^{-1}$ mode, while magnon modes in the $ab$-plane demonstrate red shifts. Ideally, one would conduct polarization analyses of these modes under pressure. However, diamond anvils are strongly birefringent under stress. 

Instead, we explore the polarization characteristics of the magnon modes at ambient, together with theoretical simulations to gain insights into their pressure evolution. While most existing Raman studies have been performed in backscattering geometries on the $ab$-plane, full access to components of the Raman tensor necessarily involves explorations along both the $ab$ and side planes. The latter is essential to understand interlayer dynamics and weak ferromagnetic stacking. 
In this geometry, we find that the magnon modes are mostly of $B_{3g}$ nature in the $D_{2h}$ setting, in contrast to the $B_{1g}$ magnons measured in the $ab$-plane geometry.  Originating from random chemical disorder and residual strains at ambient pressure, the spread of energies of the $B_{1g}$ and $B_{3g}$ magnon modes are both enhanced by a distributed pressure gradient in a high-pressure environment, faithfully reflecting the various stacking conditions along the inter-plane direction. This metastability of interlayer stacking among many similar configurations is a signature of low-dimensional quantum magnetism and our approach provides a means to parse the near degeneracy in the ground state.

\begin{figure}[!tb]
    \includegraphics[width=\columnwidth]{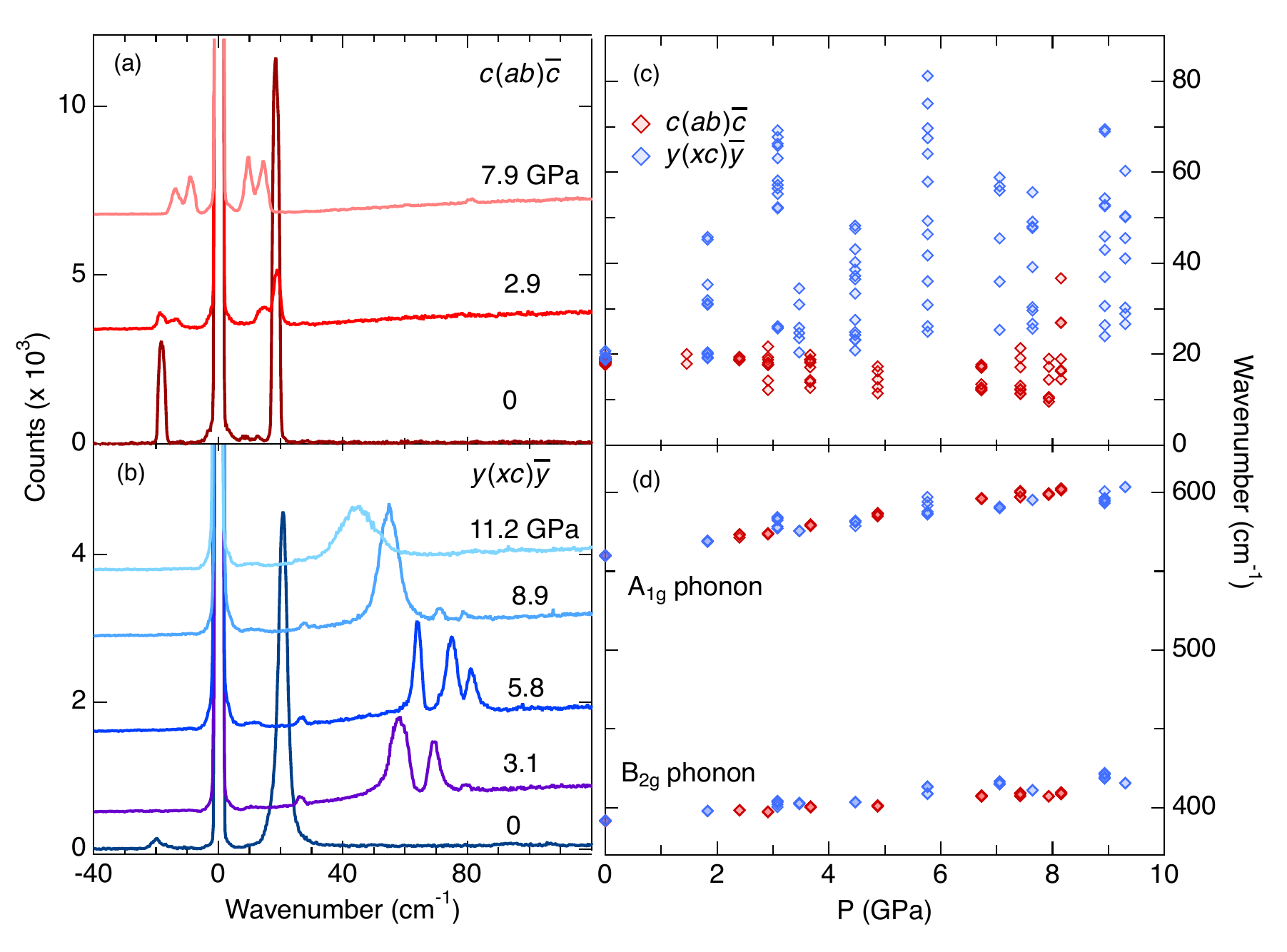}
    \caption{\label{fig:pressure}Pressure dependence of optical Raman modes in \ce{Sr2IrO4}. (left) Representative Raman spectra focusing on the low-wavenumber magnon mode, in the top-plane (a) and side-plane (b) scattering geometries for a range of pressures. The mode is weakly pressure-dependent in the top plane geometry and varies strongly with pressure in the side-plane geometry. Traces offset vertically for clarity. The ``y'' direction is along $[1\bar{1}0]$. (c)  Magnon mode energies vs pressure.  Colors indicate top vs. side plane scattering geometries. The top-plane magnon mode is red-shifted with increasing pressure and the side-plane mode is blue-shifted with significant variation in energy depending on spot locations on the sample. We attribute these variations to different local stacking configurations. (d) Representative phonon mode energies vs. pressure. In contrast to the magnon behavior, the phonon energies do not vary significantly with laser spot position and evolve approximately linearly with pressure for both scattering geometries. }
\end{figure}

\section{Results}
We plot in Fig. \ref{fig:pressure} several Raman spectra of \ce{Sr2IrO4} at T = 5 K under high pressure up to 11 GPa, in both $ab$-plane (a) and side-plane (b) geometries \cite{liOpticalRamanMeasurements2020}. With measurements over many pieces of single crystal samples at different pressures, we observe that magnon modes from the $ab$ plane are predominantly in the range between 10 and 20 cm$^{-1}$, with a few modes exhibiting energies up to 40 cm$^{-1}$ at high pressures exceeding 8 GPa. By contrast, the magnon modes from the side-plane are all in the range of 20 to 80 cm$^{-1}$. The Raman spectra also vary across the sample surface at a specific pressure, so those modes are position dependent \cite{liMagneticOrderDisorder2021}. All observed magnon modes are marked over $\omega-P$ space in Fig. \ref{fig:pressure}(c). There is clear mutual exclusion between side-plane and $ab$-plane scattering configurations. We contrast this with the behavior of the phonon modes (Fig. \ref{fig:pressure}(d)), where the mode energies do not vary significantly with position on the samples and evolve approximately linearly with pressure, indicating that pressure is not significantly altering or distorting the lattice. While the wide spread of magnon modes becomes prominent around 3 GPa, the pressure boundary of this behavior is not strict; the mode spread is observable at 2 GPa, and antecedents can exist even at ambient pressure as demonstrated by the 10 cm$^{-1}$ mode reported in the literature \cite{gimIsotropicAnisotropicRegimes2016}. At higher pressures, a spin-flip transition renders the magnon modes Raman-silent in both scattering geometries \cite{liMagneticOrderDisorder2021}, but previous magnetic x-ray experiments show that the magnitude of the spin moments remains essentially unchanged across this transition \cite{haskelPossibleQuantumParamagnetism2020}. The appearance of magnon modes at ambient pressure that are characteristic of the high-pressure response permits us to use polarization analysis to explore further the nature of these modes.

\begin{figure}[!tb]
    \centering
    \includegraphics[width=\columnwidth]{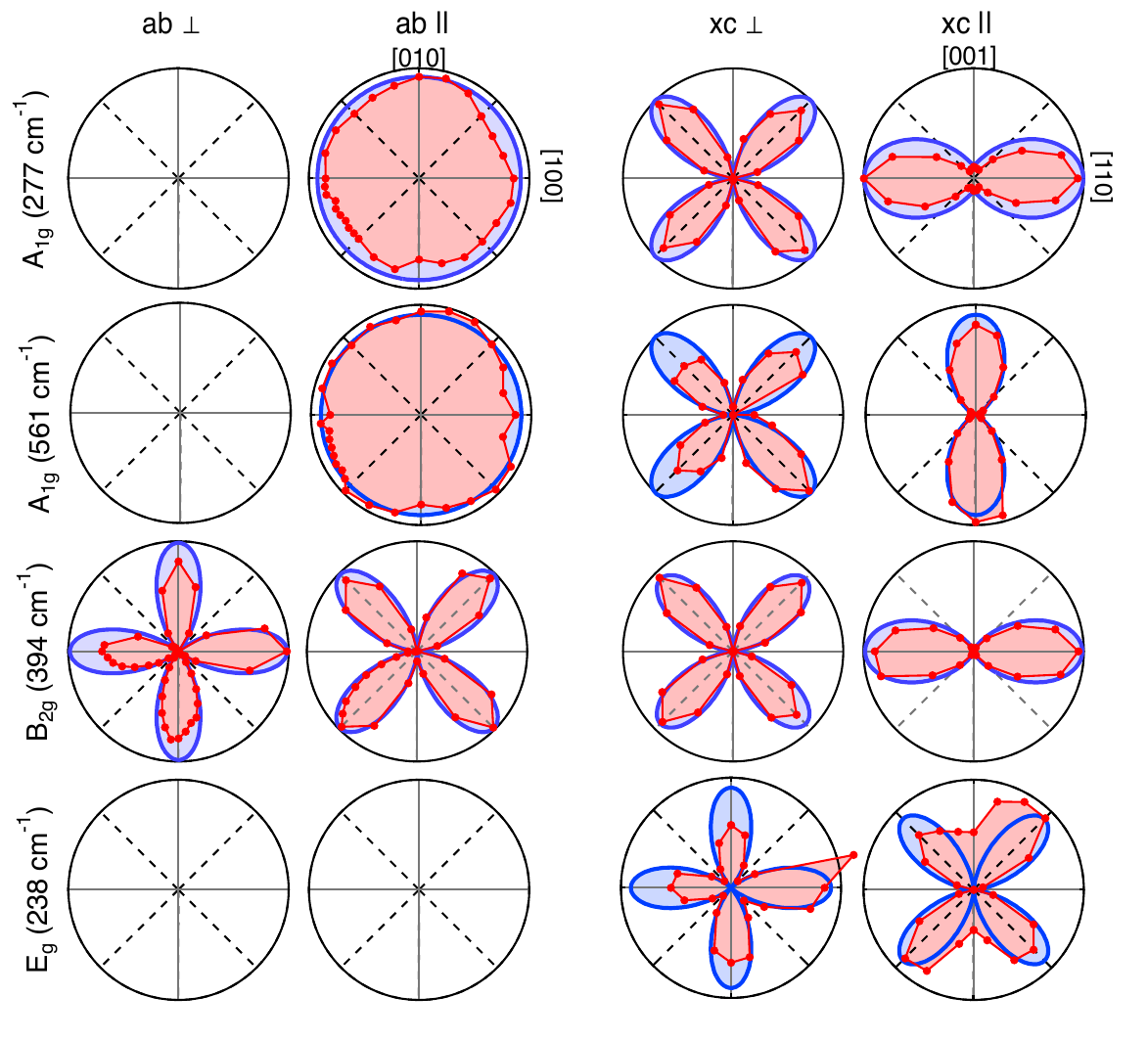}
    \caption{Angle-resolved polarization Raman spectra for four phonon modes in \ce{Sr2IrO4}. The $ab$ (top plane) and $xc$ (side plane) scattering geometries are described in the text; $||$ and $\perp$ denote polarizer and analyzer parallel and perpendicular to each other, respectively. Red points are normalized areas of the peak in the Raman spectra centered at the indicated wavenumber; blue curves are the predicted behavior of an ideal $D_{4h}$ lattice.}
    \label{fig:phonon_4mode}
\end{figure}

\begin{figure}[!tb]
    \centering
    \includegraphics[width=\columnwidth]{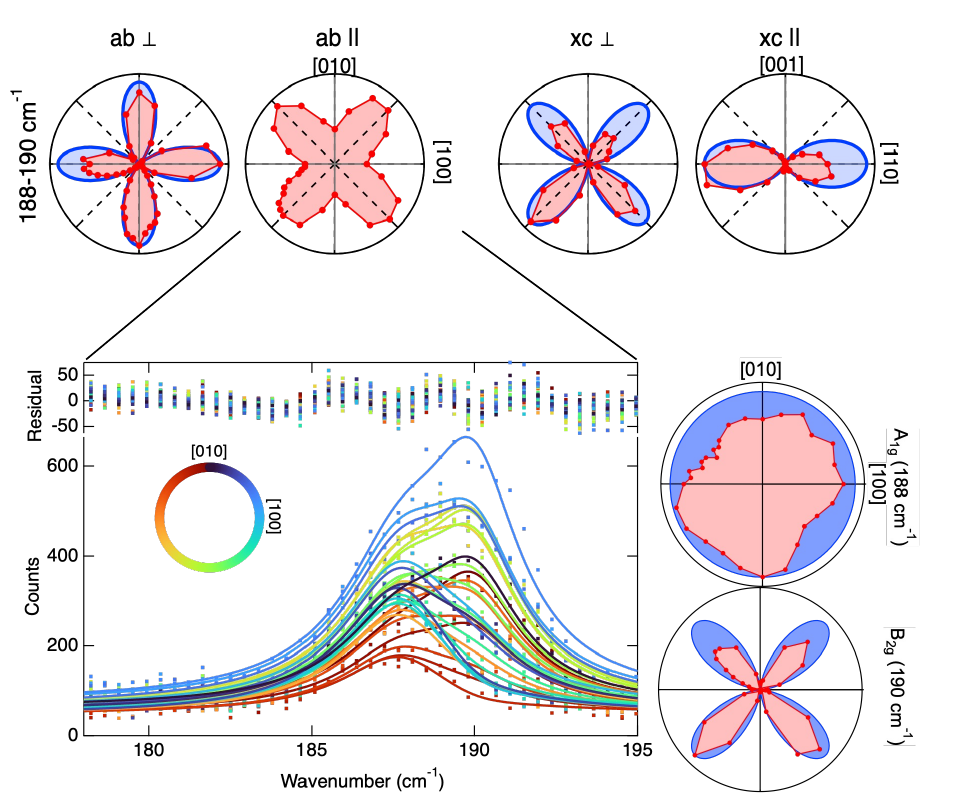}
    \caption{ Angle-resolved polarization Raman spectra for overlapping $A_{1g}$ and $B_{2g}$ phonon modes in \ce{Sr2IrO4}. (top) Integrated Raman intensity vs. angle for the four scattering geometries shown in Fig. \ref{fig:phonon_4mode}. For the $ab$ $\perp$ geometry, $A_{1g}$ is identically zero and the response is entirely due to the $B_{2g}$ mode. For both side-plane modes, $A_{1g}$ and $B_{2g}$ are predicted to have the same angular response. (bottom) For the $ab$ $||$ mode, a self-consistent two-peak global fit is performed (see text for details), allowing the clear observation of an $A_{1g}$ mode at 188 cm$^{-1}$ and a $B_{2g}$ mode at 190 cm$^{-1}$. }
    \label{fig:phonon_188}
\end{figure}

We first examine the lattice behavior. Conventionally, polarization studies use four configurations XX, X’X’, XY, X’Y’, parallel and perpendicular (crossed) incident/detection polarizers along the $[100]$ or $[110]$ directions \cite{gretarssonRamanScatteringStudy2017}. Here, we use a more in-depth angle-resolved polarization technique, measuring a series of polarizer/analyzer orientations and extracting integrated peak intensity for several modes (Figs. \ref{fig:phonon_4mode} and \ref{fig:phonon_188}). The nature of each mode is identified by comparing the polarization patterns with those expected from the Raman selection rules. For the geometries measured, $D_{4h}$ and $C_{4h}$ yield equivalent polarization patterns. We adopt here the nomenclature of the $D_{4h}$ group. The full set of patterns is collected in Fig. S1. 

At $\omega\sim190$ cm$^{-1}$, previous measurements identified two closely positioned Raman modes in \ce{Sr2IrO4}: an $A_{1g}$ mode at 188 cm$^{-1}$ and a $B_{2g}$ mode at 191 cm$^{-1}$ \cite{gretarssonRamanScatteringStudy2017}. These lines are too close in energy to be easily differentiated in an unpolarized approach \cite{liMagneticOrderDisorder2021}. Fig. \ref{fig:phonon_188} shows the angle-dependent response in this channel. For three of the four measurement geometries, interpretation is straightforward. In the top-plane geometry, $A_{1g}$ has no intensity in a crossed-polarizer configuration, so the measured response is expected to be purely due to the $B_{2g}$ mode. In the side-plane geometry, $A_{1g}$ and $B_{2g}$ modes have the same calculated angular dependence for both parallel and crossed polarizer/analyzer configurations (Fig. S1). However, the two modes have different functional forms in the top-plane parallel configuration. To robustly separate the two modes, we fit the 24 spectra at different angular positions relative to the sample using a global fitting of the whole set with common peak positions and widths but different amplitudes (Fig. \ref{fig:phonon_188} bottom). The full data set collectively reveals symmetries of these two finely separated modes, as the 187 cm$^{-1}$ mode is of nearly circular form, suggesting an $A_{1g}$ type and the 189 cm$^{-1}$ mode has a four-fold flower pattern of the $B_{2g}$ type. For the side plane geometry, we can conclude from the measured spectral shape of the 394 cm$^{-1}$ $B_{2g}$ mode that the surface is spanned by the $c$-axis and the [110] axis but not the [100] axis. From here on, we denote the side-plane of the current sample as $xc$. It has been noted in Ref. \cite{kimSingleCrystalGrowth2022} that platinum impurities included in \ce{Sr2IrO4} crystals during growth can introduce a Raman mode at 705-728 cm$^{-1}$. We use this feature to calibrate the instrumental systematics (Supplemental Fig. S3).

Our analysis has identified a total of four $A_{1g}$ modes in the $D_{4h}$ setting (188, 277, 334, and 561 cm$^{-1}$), consistent with observations reported in Ref. \cite{gretarssonRamanScatteringStudy2017}. For these modes, the lack of detectable intensities in the cross-polarization spectra of the $ab$ plane can set a limit of potential anisotropy at $< 1\%$, strongly supporting the local four-fold symmetry in the lattice. On the other hand, with the extra insight provided by the $xc$-plane geometry, we see that the two $A_{1g}$ modes of 277 and 561 cm$^{-1}$ have opposite shapes in the parallel polarization configuration of the $xc$-plane geometry. More specifically, the 561 cm$^{-1}$ mode has nodes with nearly zero intensity along the in-plane [110] direction, while the 277 cm$^{-1}$ mode has reduced intensities but no zero node along [001] (Fig.~\ref{fig:phonon_4mode}). This suggests that the $c$-axis coupling constant is weaker than the couplings in the $ab$ plane for the 277 cm$^{-1}$ mode. The converse is true for the 561 cm$^{-1}$ mode. From the literature, the 277 cm$^{-1}$ mode is attributed to bending of in-plane Ir-O-Ir bonds, while the 561 cm$^{-1}$ mode is due to oxygen motion \cite{cetinCrossoverCoherentIncoherent2012}. Explorations along the side-plane geometry also allow us to directly detect the $E_g$ modes, which do not have spectral weight in the ab-plane backscattering configuration. Here, we fully verify that the mode at 238 cm$^{-1}$ has $E_g$ symmetry, consistent with our previous study \cite{liMagneticOrderDisorder2021}. 

\begin{figure}[!tb]
    \centering
    \includegraphics[width=\columnwidth]{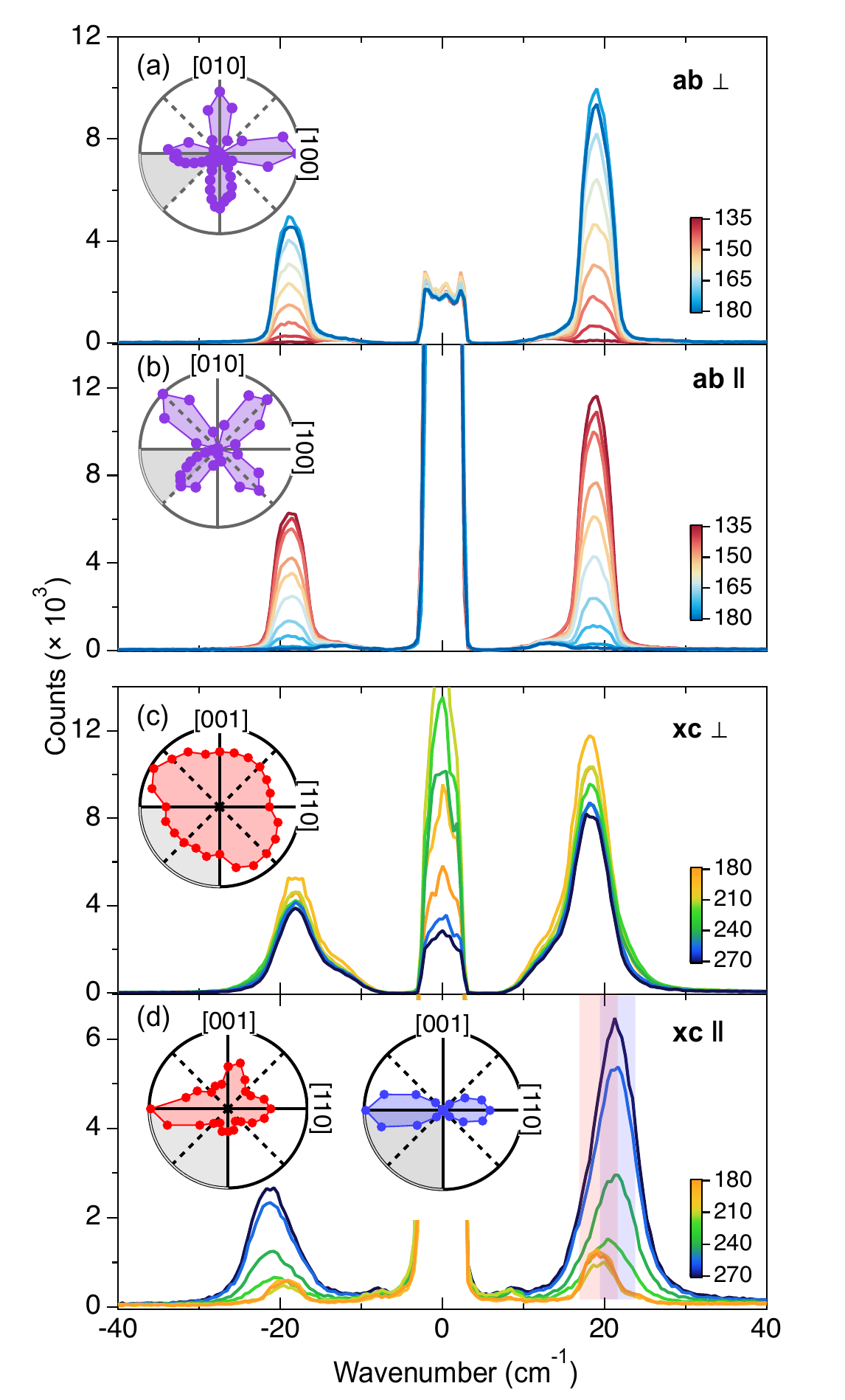}
    \caption{Angle-resolved polarization data of \ce{Sr2IrO4} magnon modes at $T=5$~K. Main panels show low-wavenumber raw spectra over one quadrant or octant of angles with trace color indicating the angle; insets show the full angular dependence of the peak area. The $xc$ $||$ scattering geometry (panel d) exhibits two overlapping modes at different frequencies; shaded rectangles denote the approximate frequencies of the two modes. See text and Fig. \ref{fig:magnon_bc} below for details on the separation procedure. }
    \label{fig:magnon}
\end{figure}

\begin{figure}[!tb]
    \centering
    \includegraphics[width=\columnwidth]{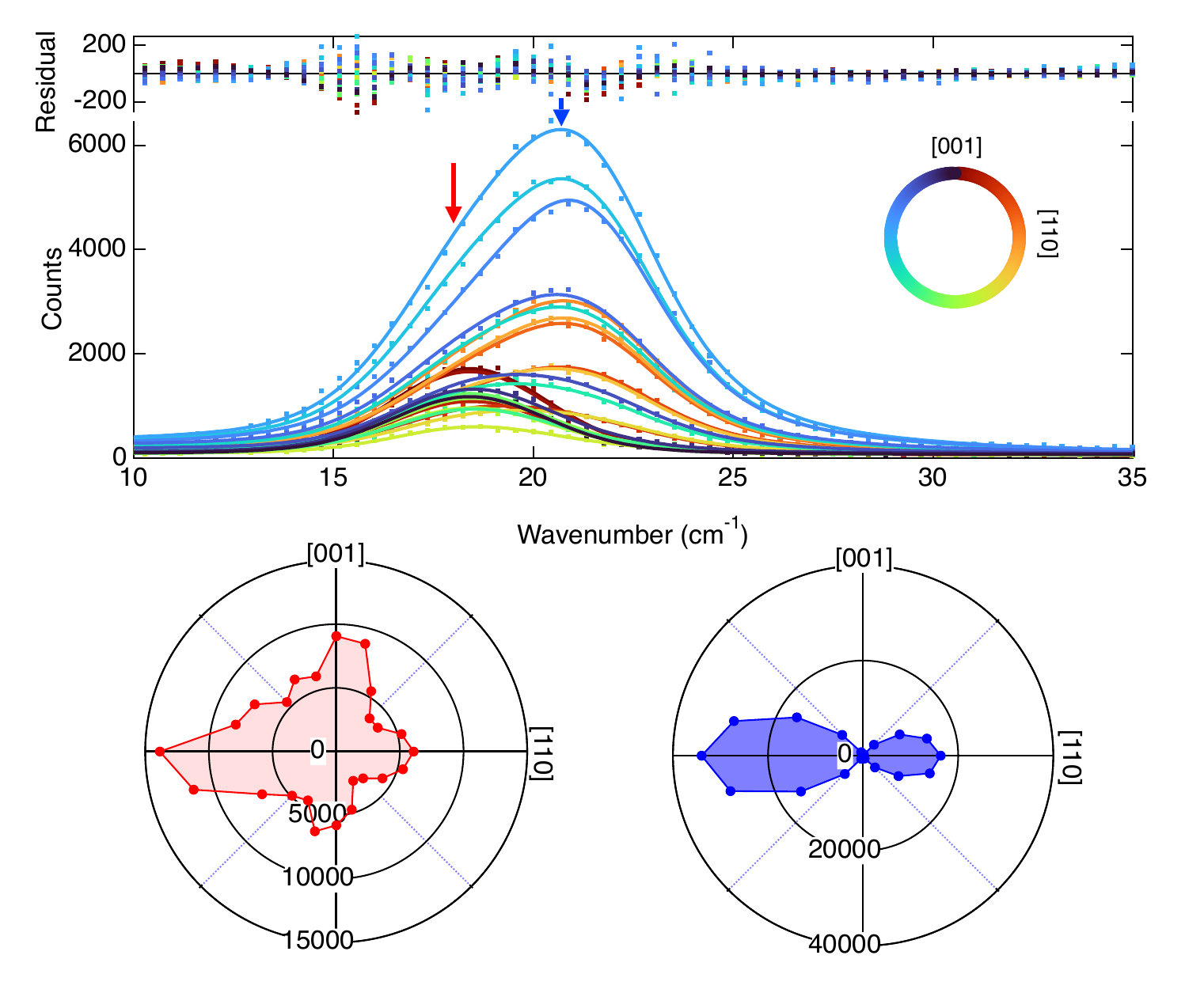}
    \caption{Angle-resolved polarization Raman spectra for overlapping magnon modes in \ce{Sr2IrO4}. Top: Raman spectra in the $xc$ (side plane) geometry with polarizer parallel to analyzer. (bottom) A self-consistent global fit to two overlapping peaks (see text for details) yields two modes with contrasting angular behavior at 18 and 21 cm$^{-1}$. }
    \label{fig:magnon_bc}
\end{figure}

With our measurement and analysis protocol established through the lattice phonon study, we now turn to the low-frequency regime for the magnon behavior (Figs. \ref{fig:magnon} and \ref{fig:magnon_bc}). In the $ab$-plane geometry, there is a single mode at 18.0 cm$^{-1}$, which displays clear four-fold polar patterns in both parallel and crossed channels, oriented at 0 and 45°, respectively (Fig. \ref{fig:magnon}a,b).  This mode has been reported in several previous $ab$-plane studies \cite{liMagneticOrderDisorder2021,gretarssonRamanScatteringStudy2017,gimIsotropicAnisotropicRegimes2016} and was assigned to a single-magnon excitation with $B_{2g}$ symmetry based on the crystallographic $D_{4h}$ point group \cite{gretarssonRamanScatteringStudy2017}. For a magnetic excitation, however, its symmetry should more properly be classified using the magnetic point group. Within the $D_{2h}$ magnetic point group \cite{yeStructureSymmetryDetermination2015}, the polarization dependence of the 18.0 cm$^{-1}$ mode is consistent with a $B_{1g}$-type excitation. For the geometries considered here, this $B_{1g}$ response produces the same polarization patterns as the $B_{2g}$ response in $D_{4h}$ \cite{gretarssonRamanScatteringStudy2017}, reconciling our assignment with the earlier literature.
The full set of calculated intensity patterns for $D_{2h}$ is shown in Fig. S2. In our crystal at ambient pressure, we did not observe the 10 cm$^{-1}$ mode (Fig. \ref{fig:magnon}) reported in Ref. \cite{gimIsotropicAnisotropicRegimes2016}. 

With the laser beam normal to the side plane, the 18 cm$^{-1}$ mode is still present as reported previously \cite{liMagneticOrderDisorder2021}. Here we revisit this mode under ambient conditions with full angle-resolved capabilities. In the $xc$ orientation, unlike the $ab$-plane, this 18 cm$^{-1}$ mode is never fully suppressed under crossed polarization, producing a circular pattern with a $\pm20\%$ variation in integrated intensity and no nodes (Fig. \ref{fig:magnon}(c)). Finally, in the $xc$-parallel configuration (Fig. \ref{fig:magnon}(d)), two separate modes at 18.4 and 21.3 cm$^{-1}$ emerge.  Following the procedure described above for separating closely-spaced phonon modes, a global fit to the entire dataset (full circle at $15^\circ$ increments, Fig. \ref{fig:magnon_bc}) is again conducted here, fitting a pair of Voigt peaks to each spectrum. The wavenumber, peak width, and Voigt shape for both modes are fit in common across all polarization orientations, and the areas of the peaks are varied independently. As shown by the fit residual in Fig. \ref{fig:magnon_bc}, this procedure yields a good match to the data while minimizing the number of free parameters in the model. The fit results for these areas are plotted in measurement units in Fig. \ref{fig:magnon_bc}, with normalized patterns for each mode shown in the insets of Fig. \ref{fig:magnon}(d). 

\section{Discussion}
Recently, polarization-specific optical Raman scattering was used to support a claim of a nematic spin phase in \ce{Sr2IrO4} \cite{kimQuantumSpinNematic2024}. Based on three cross or parallel polarization conditions of ZZ, ZX, and XY, three peaks at 2.4, 2.3, and 2.2 meV (19.3, 18.5, and 17.7 cm$^{-1}$) were attributed to $A_{1g}$, $E_g$, and $B_{2g}$ symmetries, respectively, within a $D_{4h}$-based symmetry framework.  The $A_{1g}$ peak was interpreted as a phase mode associated with quadrupolar order. The peak observed in the ZX channel was assigned $E_g$ symmetry and referred to as the quadrupolar order parameter, while the $B_{2g}$ peak was attributed to a single-magnon excitation \cite{kimQuantumSpinNematic2024}. Our study has demonstrated that polarization studies under both the $ab||$ and $ab\perp$ configurations (Figs. \ref{fig:magnon}(a,b)) always have four zero nodes, inconsistent with an $A_{1g}$ (or $A_g$ in the $D_{2h}$ setting) attribution, as these modes would demand an isotropic (or nearly isotropic) form under $ab\:||$ and vanishing values under $ab\perp$ (Figs. S1 and S2). The polarization dependence of the 18 cm$^{-1}$ mode has complex, non-vanishing intensities along every direction in both $xc\perp$ and $xc\:||$ configurations (Figs. \ref{fig:magnon}(c,d)). Combined, this leaves the 18 cm$^{-1}$ mode as a combination of $B_{1g}$ and $B_{3g}$ symmetries in the $D_{2h}$ setting (or $B_{2g}$ and $E_g$ in the $D_{4h}$ setting in Ref. \cite{kimQuantumSpinNematic2024}). Our phonon studies in Fig. \ref{fig:phonon_4mode} provide an example of such a superposition. The non-vanishing intensity of the $E_g$ phonon along the [0, 0, 1] direction (Fig. \ref{fig:phonon_4mode}d) under the $xc||$ configuration explains the all-around non-vanishing intensity of the 18 cm$^{-1}$ mode in Fig. \ref{fig:magnon}(d). This subtlety of the $E_g/B_{3g}$ forms highlights the importance of a full polarization study of the Raman mode in both configurations. The polarization dependence of the low-energy Raman response does not require the additional quadrupolar order proposed in Ref. \cite{kimQuantumSpinNematic2024}, but can be accounted for by conventional magnon excitations within the established magnetic symmetry.



All three assigned peaks in Ref. \cite{kimQuantumSpinNematic2024}, with a FWHM of 0.4-0.6 meV at the base temperature of 30 K, are within a separation of half width at half maximum to each other. One potential explanation of the small differences in peak center is sample inhomogeneity, a known chemistry challenge of iridate materials in general \cite{vlaskovaMagneticPropertiesCrystal2020} even without issues introduced by Pt impurities \cite{kimSingleCrystalGrowth2022}. Our previous study at ambient pressure demonstrated that the peak energy of the 18 cm$^{-1}$ mode can have a small fluctuation based on the location of the probe across the sample surface (Fig. 2b of Ref. \cite{liMagneticOrderDisorder2021}). The peak center differences in Ref. \cite{kimQuantumSpinNematic2024} are also consistent with our observed peak spread between 18.4 and 21.3 cm$^{-1}$. We believe that the 21.3 cm$^{-1}$ mode is most likely an antecedent of the high-pressure modes in Fig. \ref{fig:pressure}, with its origin arising from defects and chemical disorder in the ambient pressure sample. Nevertheless, its symmetry is close to a $B_{2g}$ or $B_{3g}$ mode in the $D_{2h}$ point group (or $E_g$ in $D_{4h}$) with no $ab$-plane spectral weight, which explains the contrasting blue and red shifts of the modes in Fig. \ref{fig:pressure}(c). The small admixture of symmetry-distinct Raman components in the side-plane polarization dependence indicates a weak departure from the ideal $D_{2h}$ selection rules, potentially reflecting the subtle, and possibly sample-dependent, symmetry lowering reported in \ce{Sr2IrO4} \cite{kimSingleCrystalGrowth2022,yeMagneticCrystalStructures2013,torchinskyStructuralDistortionInducedMagnetoelastic2015}. 

Carter et al. \cite{carterTheoryMetalinsulatorTransition2013} discussed the charge gap in \ce{Sr2IrO4} as determined by a Hubbard $U$ of $\sim0.46$ eV, typical of $5d$ systems. They also pointed out that antiferromagnetism is formed due to couplings across the next-nearest-layer distance, of much smaller energies, to build the up-up-down-down ($uudd$) stacking of Ir layers in one unit cell. More quantitative values of magnetism are provided in other studies. Ref. \cite{takayamaModelAnalysisMagnetic2016}  specifies the leading in-plane exchange coupling $J_{ab}\sim 0.1$ eV that is comparable to the Neel temperature $T_N \sim 230$ K or $\sim21$ meV, and a leading interlayer antiferromagnetic coupling $J_c \sim 16$ $\mu$eV. On the other hand, based on the canted ferromagnetic moment ($\sim0.07 \mu_B$/Ir) and an in-plane coercive field ($\sim0.15$ T), Ref. \cite{takayamaModelAnalysisMagnetic2016} estimated the extra energy associated with the $c$-axis stacking as $\Delta\sim0.7$ $\mu$eV ($\sim8$ mK), equivalent to the range of near degeneracy in the meta-magnetism associated with various $c$-axis stackings. This set of diverse energy scales over nearly six decades is consistent with later estimations in Ref.   \cite{porrasPseudospinlatticeCouplingSpinorbit2019}, with in-plane vs. out-of-plane main coupling ratio $J_{ab}/J_c > 3000$ and anisotropic interlayer coupling $\Delta=0.02 J_c$. Such a large range of energy scales highlights the general difficulty of strongly correlated electron systems and explains the challenge to properly understand the interlayer coupling in \ce{Sr2IrO4}. 

Having presented our experimental observations, we now turn to theory to explicate the physical origins of the magnon modes. At the $\Gamma$ point, magnon selection rules are set by the unitary invariant subgroup of the magnetic point group, instead of the lattice point group $D_{4h}$. In \ce{Sr2IrO4}, the magnetic order reduces the symmetry to the magnetic point group $mmm.1^\prime$, as the antiferromagnetism breaks the in-plane four-fold symmetry resulting in the unitary invariant subgroup $mmm\; (D_{2h})$. The classical ground state is the canted Néel state, shown in Supplementary Figure 4: the staggered moment lies along the crystallographic $a$ axis, pinned there by an intralayer anisotropy $\Gamma_1$ (discussed below), and the DM canting tilts it toward a small net moment along $b$, $\langle S \rangle=S(\cos\phi \;\hat{a} \pm \sin\phi \; \hat{b} )$, with canting angle fixed by the DM/exchange ratio, $\tan2\phi=\;D/J_1  \; (\phi\sim13^\circ)$  with the four layers stack in the $uudd$ net-moment sequence. The \ce{Ir^4+} moments ordering structure is captured by the $8c$ Wyckoff positions of magnetic space group MSG \textit{PIcca} (54.352). From the magnon band representation induced by these Wyckoff positions of this MSG and restricted to the $\Gamma$ point, all the magnon bands at the Brillouin zone center are assigned one-dimensional representations \cite{MTQCElcoro_2021,Cata4444Xu_2020,karaki2024highthroughputsearchtopologicalmagnon}:
\begin{equation}
\begin{aligned}
(A_{\mathrm{Ir}})_{8c}\uparrow P_{Icca}(54.352)\downarrow\Gamma
={}& \Gamma_{1}^{+}(1)\oplus\Gamma_{1}^{-}(1) \\
&\oplus\Gamma_{2}^{+}(1)\oplus\Gamma_{2}^{-}(1) \\
&\oplus\Gamma_{3}^{+}(1)\oplus\Gamma_{3}^{-}(1) \\
&\oplus\Gamma_{4}^{+}(1)\oplus\Gamma_{4}^{-}(1).
\end{aligned}
\end{equation}

This is fully consistent with the constraints of the unitary subgroup of the magnetic point group $D_{2h}$ symmetry, which only permits non-degenerate irreps, with the Raman active ones $A_g (\Gamma_1^{+})$, $B_{1g} (\Gamma_3^{+})$, $B_{2g} (\Gamma_2^{+})$ and $B_{3g} (\Gamma_4^{+})$.\\

To further investigate the magnons, we carried out linear spin-wave (LSW) calculations starting from the full spin Hamiltonian, adapted from Porras \textit{et al.}~\cite{porrasPseudospinlatticeCouplingSpinorbit2019}, which was obtained by fitting to RIXS and magnetization data:
\begin{equation}
H_{\mathrm{total}}
=
H_{\mathrm{iso}}
+
H_{\mathrm{ani}}^{1}
+
H_{\mathrm{ani}}^{2}
+
H_{\mathrm{aniso}}^{\mathrm{sp-lat}} .
\end{equation}
Here, $H_{\mathrm{iso}}$ is the isotropic Heisenberg exchange part,
\begin{equation}
H_{\mathrm{iso}}
=
\sum_{\langle ij\rangle}
J_{ij}\,\vec{S}_{i}\cdot\vec{S}_{j}
+
J_{1c}\,\vec{S}_{i}\cdot\vec{S}_{j}
+
J_{2c}\,\vec{S}_{i}\cdot\vec{S}_{j},
\end{equation}
where $J_{ij}$ runs over the first ($J_{1}$), second ($J_{2}$), and third ($J_{3}$) nearest neighbors within the $ab$ plane, and $J_{1c}$ and $J_{2c}$ are the first- and second-nearest interlayer interactions. The tetragonal distortion and rotation of the oxygen octahedra contribute Ising and Dzyaloshinskii-Moriya (DM) interaction anisotropies of the form
\begin{equation}
H_{\mathrm{ani}}^{1}
=
\sum_{\langle ij\rangle}
J_{z}S_{i}^{z}S_{j}^{z}
+
\vec{D}\cdot
\left(
\vec{S}_{i}\times\vec{S}_{j}
\right),
\end{equation}
which confine the spins to the $ab$ plane, resulting in an out-of-plane magnon gap, and give rise to a canting angle of approximately $13^{\circ}$. An interlayer anisotropy over the next-nearest-neighbor interlayer bonds takes the form
\begin{equation}
H_{\mathrm{ani}}^{2}
=
\sum_{\langle ij\rangle}
\pm\Delta_{c}
\left(
S_{i}^{a}S_{j}^{a}
-
S_{i}^{b}S_{j}^{b}
\right),
\end{equation}
where the sign depends on the bond~\cite{porrasPseudospinlatticeCouplingSpinorbit2019}. This anisotropy was found to lift the degeneracy between the \textit{uudd} and \textit{uddu} stacking orders and stabilize the observed magnetic structure of \textit{uudd} in the domain where the magnetic moments are mostly along the $a$ axis and \textit{lrrl} in the domain where the moments are mostly along the $b$ axis. The last anisotropy is an in-plane spin--lattice coupling:
\begin{equation}
\begin{aligned}
H_{\mathrm{sp-lat}}
={}&
\sum_{\langle ij\rangle}
\Gamma_{1}\cos(2\theta)
\left(
S_{i}^{x}S_{j}^{y}+S_{i}^{y}S_{j}^{x}
\right)
\nonumber\\
&-
\Gamma_{2}\sin(2\theta)
\left(
S_{i}^{x}S_{j}^{x}-S_{i}^{y}S_{j}^{y}
\right).
\end{aligned}
\end{equation}
Here, $\theta$ is the angle between the canted ferromagnetic moments and the $a$ axis. This term introduces an in-plane anisotropy and binds the magnetic moments along either the $a$ or $b$ axis. Additional details on the LSW calculation are given in the Supplemental Information. 

\begin{figure}
    \centering
    \includegraphics[width=\columnwidth]{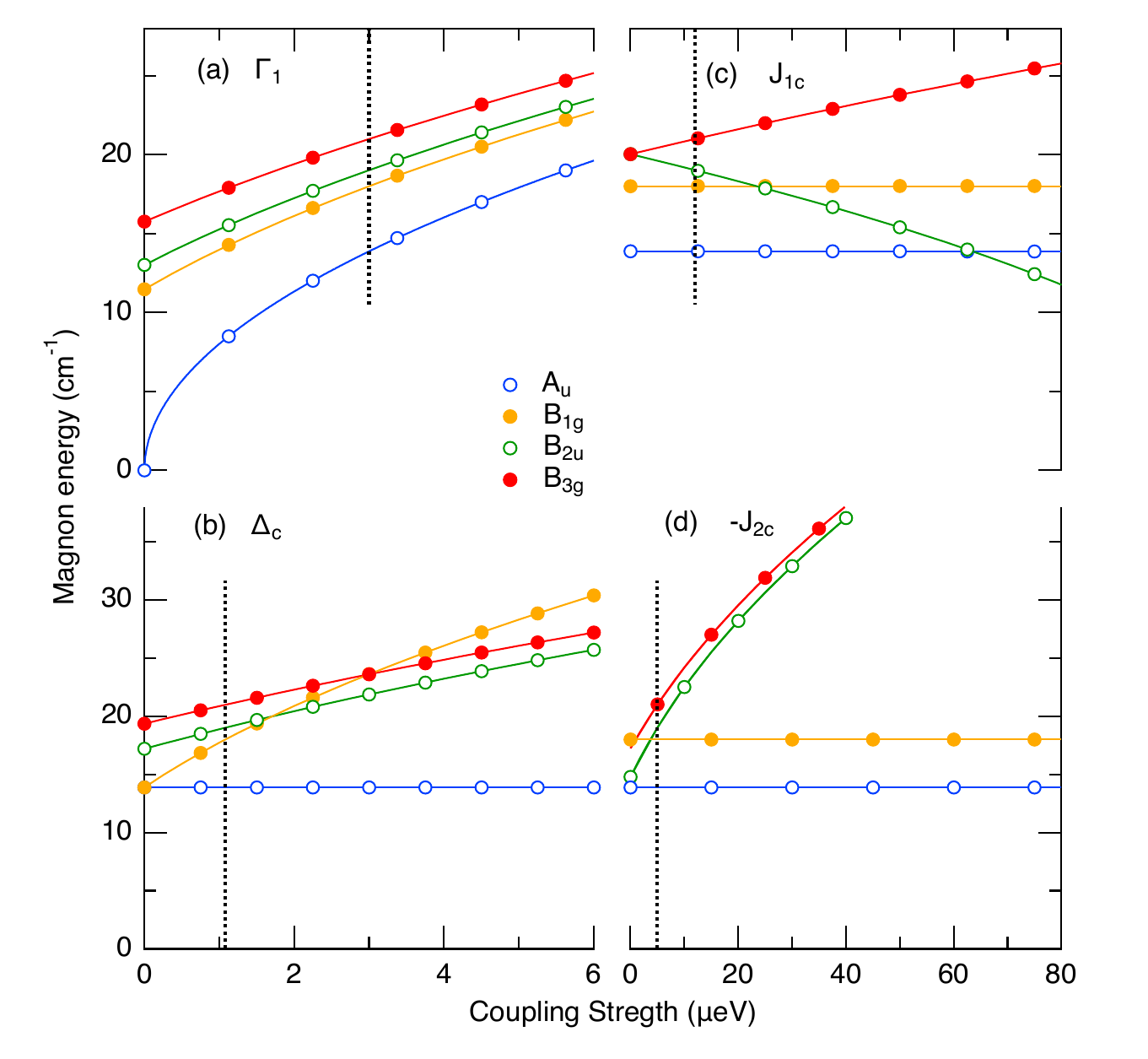}
    \caption{Dependence of the low energy magnons of \ce{Sr2IrO4} on (a) the pseudospin-lattice coupling $\Gamma_1$, (b) interlayer anisotropy $\Delta_c$, (c) interlayer coupling $J_{1c}$, and (d) second-nearest-layer coupling $J_{2c}$. For each panel, the other model parameters are the modified parameters given in Table~\ref{tab:spin_parameters} and the vertical dashed line marks the best-fit value of the coupling constant. Filled symbols indicate Raman-active modes; open symbols indicate Raman-silent modes.}
    \label{fig:calculation}
\end{figure}

Our calculations reveal a total of eight magnon modes corresponding to the eight $\mathrm{Ir}^{4+}$ spins in a unit cell. The calculated dispersions are shown in Supplementary Figure 5; here, we focus on the properties at the $\Gamma$ point. Four high-energy modes are nearly degenerate at approximately $40~\mathrm{meV}$, closely matching the out-of-plane magnon gap reported in the resonant inelastic x-ray scattering (RIXS) experiment of Porras \textit{et al.}~\cite{porrasPseudospinlatticeCouplingSpinorbit2019}. Among the low-energy modes, we identify two even-parity Raman-active modes with $B_{1g}$ and $B_{3g}$ symmetries, and two odd-parity Raman-silent modes with $A_u$ and $B_{2u}$ symmetries. Calculating the energies of these modes using the interlayer coupling constants given in Ref.~\cite{porrasPseudospinlatticeCouplingSpinorbit2019} (Table~\ref{tab:spin_parameters} left) underestimates the energies seen in Raman, both here and previously reported~\cite{liMagneticOrderDisorder2021,gretarssonRamanScatteringStudy2017,gimIsotropicAnisotropicRegimes2016}. It is important to note, however, that those parameters were obtained by fitting magnetization data measured under different applied magnetic fields and that the results of these fits are not unique. We use the additional constraints of our measured $B_{1g}$ mode at $18~\mathrm{cm}^{-1}$ and $B_{2g}$ or $B_{3g}$ mode at $21~\mathrm{cm}^{-1}$ to refine the fit for the interlayer couplings, yielding the values in the right column of Table~\ref{tab:spin_parameters}. The other two modes are Raman silent: a $B_{2u}$ mode at $2.21~\mathrm{meV}$ ($17.8~\mathrm{cm}^{-1}$) and an $A_{u}$ mode at $1.63~\mathrm{meV}$ ($13.2~\mathrm{cm}^{-1}$).  The dependences of the energies of all four modes on the interlayer couplings are shown in Fig.~\ref{fig:calculation}; the best-fit values of the couplings are indicated by dashed lines. 

\begin{table}[t]

\begin{ruledtabular}
\begin{tabular}{lcc}
Parameter
& Porras \textit{et al.}~\cite{porrasPseudospinlatticeCouplingSpinorbit2019}
& This work \\
\hline
$J_{1}$ (meV)       & $57.0$  & $57.0$  \\
$J_{2}$ (meV)       & $-16.5$ & $-16.5$ \\
$J_{3}$ (meV)       & $12.4$  & $12.4$  \\
$J_{z}$ (meV)       & $2.9$   & $2.9$   \\
$D$ (meV)           & $28.0$  & $28.0$  \\
\hline
$\Gamma_{1}$ ($\mu$eV) & $2.7$   & $3.0$  \\
$J_{1c}$ ($\mu$eV)     & $16.4$  & $12.0$  \\
$J_{2c}$ ($\mu$eV)     & $-6.2$  & $-5.0$ \\
$\Delta_{c}$ ($\mu$eV) & $0.3$  & $1.1$  \\
\end{tabular}
\end{ruledtabular}
\caption{\label{tab:spin_parameters}
Spin-model parameters. The in-plane exchange parameters are determined from RIXS measurements~\cite{pinciniAnisotropicExchangeSpinwave2017,GretarssonPhysRevLett.117.107001} and are held fixed throughout. The four interlayer and anisotropy couplings determine the low-energy $\Gamma$ quartet, see Fig. \ref{fig:calculation}. Starting from the values reported by Porras \textit{et al.}~\cite{porrasPseudospinlatticeCouplingSpinorbit2019}, they are refined here using the measured Raman gaps ($B_{1g}=18~\mathrm{cm}^{-1}$ and $B_{3g}=21~\mathrm{cm}^{-1}$), resulting in a modest shift toward a larger interlayer anisotropy $\Delta_c$ and slightly weaker interlayer exchange energies. }
\end{table}

In contrast to the small lattice energy differences associated with various $c$-axis stacking configurations~\cite{takayamaModelAnalysisMagnetic2016}, the relative energy spread of the magnon modes is much larger. At ambient pressure, the difference between the two modes is approximately $3~\mathrm{cm}^{-1}$ ($\sim 0.37~\mathrm{meV}$). Under pressure, the magnon modes spread from $20$ to above $60~\mathrm{cm}^{-1}$ ($2.5$--$7.5~\mathrm{meV}$), as shown in Fig. \ref{fig:pressure}. Both the energy spread and the symmetries of the $B_{3g}$ and $B_{1g}$ modes in our experimental study (Fig. \ref{fig:pressure}) can be well understood by theoretical calculations based on microscopic insights as follows.  Quantitatively, a pressure coordinate can be understood as a parameter to tune the interlayer couplings to approximately reproduce the experimental behavior. We thus vary the interlayer couplings and anisotropy terms around their ambient-pressure values to examine how the four low-energy magnon modes respond to individual microscopic parameters (Fig. \ref{fig:calculation}). The calculations reveal a clear contrast between the two Raman-active modes: the $B_{3g}$ side-plane magnon is strongly sensitive to both $J_{1c}$  and $J_{2c}$,  whereas the $B_{1g}$ in-plane magnon is nearly insensitive to these interlayer exchange interactions. Because \ce{Sr2IrO4} is quasi-two-dimensional, pressure is expected to reduce the interlayer spacing more readily than it changes the \ce{IrO6} octahedral rotations or distortions, thereby enhancing the interlayer couplings and naturally favoring a hardening of the side-plane mode. This tendency may remain relevant even above the pressure-induced advent of mixed antiferromagnetic stackings near 2–3 GPa. In contrast, the $B_{1g}$ mode responds more strongly to the anisotropy terms, particularly the pseudospin-lattice anisotropy $\Gamma_1$ and, notably, the interlayer anisotropy $\Delta_c$. The continuous parameter variations in Fig. \ref{fig:calculation} therefore should not be interpreted as a direct simulation of the experimentally observed transition, since pressure modifies several interactions simultaneously and may change the energetically favored stacking order. The near degeneracy of competing stacking configurations may also contribute to the spatial variation of the magnon energies observed at a given pressure in Fig. \ref{fig:pressure}. Instead, the calculations clarify the distinct microscopic sensitivities of the in-plane and side-plane magnons and provide a qualitative framework for understanding their different energies, pressure responses, and spatial variations observed in Fig. \ref{fig:pressure}.

 The presence of lattice defects based on either chemical disorder or pressure gradient induced stress can cause significant differences in the local interlayer couplings and anisotropies. In turn, the spin stacking along the $c$-axis can vary between possible forms such as $duud$, $udud$, and $uddu$  \cite{liMagneticOrderDisorder2021,porrasPseudospinlatticeCouplingSpinorbit2019}. The Raman magnon modes corresponding to each spin configuration are thus independent and can spread over a large range of energy.  With increasing pressure, it has been well documented that there exists an increasing amount of local pressure variation  \cite{Feng:2010dg}. We thus expect the increase of both the energy and the spread of magnon energy in Fig. 1. The emergence and blue shift of many different Raman modes in the $ac$- or $xc$-configurations under pressure and their absence in an ideal antiferromagnet without strain fields or defects then can be understood.

\section{Conclusions}

Despite the static nature of disorder and pressure gradients in enhancing locally preferred magnetic layer stackings at different sample positions, hydrostatic pressure in combination with an optical local probe offers an experimental pathway to lifting and revealing the ground state degeneracy (or near degeneracy). It is also fortunate that the magnon evolution under pressure has its antecedents detectable at ambient conditions, enabling this detailed study of the energies and symmetries. Our approach can be extended to other members of the Ruddlesden-Popper series \ce{Sr_{n+1}Ir_{n}O_{3n+1}} \cite{carterTheoryMetalinsulatorTransition2013}, such as \ce{Sr3Ir2O7} and \ce{SrIrO3}, and may bear on the possible emergence of a quantum spin liquid at sufficiently high pressures to suppress the antiferromagnetic order \cite{haskelPossibleQuantumParamagnetism2020}.

Our results highlight a general aspect of layered quantum magnets, in which lattice symmetry and stacking can frustrate the leading exchange pathways, leaving the magnetic ground state controlled by much weaker couplings or anisotropies. This sensitivity has been highlighted across a range of layered magnetic systems. In van der Waals magnets such as \ce{CrI3}, the magnetic order depends critically on stacking geometry, with relatively small structural changes sufficient to switch the alignment between neighboring layers \cite{songSwitching2DMagnetic2019,liPressurecontrolledInterlayerMagnetism2019}. In the Kitaev-candidate \ce{\alpha-RuCl3}, weak out-of-plane interactions and magnetoelastic couplings strongly influence the magnetic phase diagram in proximity to frustrated and spin-liquid-like regimes  \cite{stahlPressuretuningARuCl3Quantum2024}. Similarly, in the intrinsic magnetic topological insulator \ce{MnBi2Te4} \cite{padmanabhanInterlayerMagnetophononicCoupling2022} and related compounds  [11], exchange between layers is crucial to both magnetic order and topology and is strongly affected by lattice distortions and stacking. Across these systems, a unifying theme emerges: magnetic ground-state selection is not always determined by the largest interactions alone but can hinge on weaker couplings that become operative when leading terms cancel or compete.\\

\begin{acknowledgments}
Work at Caltech supported by the US Air Force Office of Scientific Research, Grant FA9550-25-1-0017. Work at the University of Toronto supported by Natural Sciences and Engineering Research Council of Canada (NSERC) Alliance International Catalyst Quantum grant ALLRP 592597-23. Y.F. acknowledges financial support from the Okinawa Institute of Science and Technology Graduate University, with subsidy funding from the Cabinet Office, Government of Japan. A.F., and Y-M. L. acknowledge funding support from the Center for Emergent Materials, an NSF MRSEC, under award number DMR-2011876.

\end{acknowledgments}

\bibliography{Sr2IrO4_magnon}

@PREAMBLE{
 "\providecommand{\noopsort}[1]{}" 
 # "\providecommand{\singleletter}[1]{#1}%" 
}

@article{Cao:2016ep,
  title = {Low-Temperature Crystal and Magnetic Structure of {$\alpha$}-{{RuCl}}{$_{3}$}},
  author = {Cao, H B and Banerjee, A and Yan, J.-Q. and Bridges, C A and Lumsden, M D and Mandrus, D G and Tennant, D A and Chakoumakos, B C and Nagler, S E},
  year = 2016,
  month = apr,
  journal = {Physical Review B},
  volume = {93},
  number = {13},
  pages = {134423},
  publisher = {American Physical Society},
  doi = {10.1103/PhysRevB.93.134423},
  langid = {english},
  uri = {papers3://publication/doi/10.1103/PhysRevB.93.134423}
}

@article{carterTheoryMetalinsulatorTransition2013,
  title = {Theory of Metal-Insulator Transition in the Family of Perovskite Iridium Oxides},
  author = {Carter, Jean-Michel and Shankar V., Vijay and Kee, Hae-Young},
  year = 2013,
  month = jul,
  journal = {Physical Review B},
  volume = {88},
  number = {3},
  pages = {035111},
  publisher = {American Physical Society},
  doi = {10.1103/PhysRevB.88.035111},
  urldate = {2025-01-13}
}

@article{cetinCrossoverCoherentIncoherent2012,
  title = {Crossover from Coherent to Incoherent Scattering in Spin-Orbit Dominated {{Sr}}{\textsubscript{2}}{{IrO}}{\textsubscript{4}}},
  author = {Cetin, Mehmet Fatih and Lemmens, Peter and Gnezdilov, Vladimir and Wulferding, Dirk and Menzel, Dirk and Takayama, Tomohiro and Ohashi, Kei and Takagi, Hidenori},
  year = 2012,
  month = may,
  journal = {Physical Review B},
  volume = {85},
  number = {19},
  pages = {195148},
  publisher = {American Physical Society},
  doi = {10.1103/PhysRevB.85.195148},
  urldate = {2021-10-05}
}

@article{chenPersistentInsulatingState2020,
  title = {Persistent Insulating State at Megabar Pressures in Strongly Spin-Orbit Coupled {{Sr}}{\textsubscript{2}}{{IrO}}{\textsubscript{4}}},
  author = {Chen, Chunhua and Zhou, Yonghui and Chen, Xuliang and Han, Tao and An, Chao and Zhou, Ying and Yuan, Yifang and Zhang, Bowen and Wang, Shuyang and Zhang, Ranran and Zhang, Lili and Zhang, Changjin and Yang, Zhaorong and DeLong, Lance E. and Cao, Gang},
  year = 2020,
  month = apr,
  journal = {Physical Review B},
  volume = {101},
  number = {14},
  pages = {144102},
  publisher = {American Physical Society},
  doi = {10.1103/PhysRevB.101.144102},
  urldate = {2021-10-05}
}

@article{choiLightinducedInsulatorMetal2024,
  title = {Light-Induced Insulator--Metal Transition in {{Sr}}{\textsubscript{2}}{{IrO}}{\textsubscript{4}} Reveals the Nature of the Insulating Ground State},
  author = {Choi, Dongsung and Yue, Changming and Azoury, Doron and Porter, Zachary and Chen, Jiyu and Petocchi, Francesco and Baldini, Edoardo and Lv, Baiqing and Mogi, Masataka and Su, Yifan and Wilson, Stephen D. and Eckstein, Martin and Werner, Philipp and Gedik, Nuh},
  year = 2024,
  month = jul,
  journal = {Proceedings of the National Academy of Sciences},
  volume = {121},
  number = {29},
  pages = {e2323013121},
  publisher = {Proceedings of the National Academy of Sciences},
  doi = {10.1073/pnas.2323013121},
  urldate = {2024-07-16}
}

@article{chubukovOrderDisorderKagome1992,
  title = {Order from Disorder in a {{Kagom\'e}} Antiferromagnet},
  author = {Chubukov, Andrey},
  year = 1992,
  month = aug,
  journal = {Physical Review Letters},
  volume = {69},
  number = {5},
  pages = {832--835},
  publisher = {American Physical Society},
  doi = {10.1103/PhysRevLett.69.832},
  urldate = {2026-06-03}
}

@article{crawfordStructuralMagneticStudies1994,
  title = {Structural and Magnetic Studies of {{Sr}}{\textsubscript{2}}{{IrO}}{\textsubscript{4}}},
  author = {Crawford, M. K. and Subramanian, M. A. and Harlow, R. L. and {Fernandez-Baca}, J. A. and Wang, Z. R. and Johnston, D. C.},
  year = 1994,
  month = apr,
  journal = {Physical Review B},
  volume = {49},
  number = {13},
  pages = {9198--9201},
  publisher = {American Physical Society},
  doi = {10.1103/PhysRevB.49.9198},
  urldate = {2026-06-03}
}

@article{Feng:2010dg,
  title = {Invited {{Article}}: {{High-pressure}} Techniques for Condensed Matter Physics at Low Temperature},
  author = {Feng, Yejun and Jaramillo, Rafael and Wang, Jiyang and Ren, Yang and Rosenbaum, Thomas F},
  year = 2010,
  journal = {Review of Scientific Instruments},
  volume = {81},
  number = {4},
  pages = {041301},
  publisher = {American Institute of Physics},
  doi = {10.1063/1.3400212},
  langid = {english},
  uri = {papers3://publication/doi/10.1063/1.3400212}
}

@article{gimIsotropicAnisotropicRegimes2016,
  title = {Isotropic and Anisotropic Regimes of the Field-Dependent Spin Dynamics in {{Sr}}{\textsubscript{2}}{{IrO}}{\textsubscript{4}} : {{Raman}} Scattering Studies},
  shorttitle = {Isotropic and Anisotropic Regimes of the Field-Dependent Spin Dynamics in {{Sr}} 2 {{IrO}} 4},
  author = {Gim, Y. and Sethi, A. and Zhao, Q. and Mitchell, J. F. and Cao, G. and Cooper, S. L.},
  year = 2016,
  month = jan,
  journal = {Physical Review B},
  volume = {93},
  number = {2},
  pages = {024405},
  issn = {2469-9950, 2469-9969},
  doi = {10.1103/PhysRevB.93.024405},
  urldate = {2019-12-12},
  langid = {english}
}

@article{gretarssonRamanScatteringStudy2017,
  title = {Raman Scattering Study of Vibrational and Magnetic Excitations in {{Sr}}{\textsubscript{2-x}}{{La}}{\textsubscript{x}}{{IrO}}{\textsubscript{4}}},
  author = {Gretarsson, H. and Sauceda, J. and Sung, N. H. and H{\"o}ppner, M. and Minola, M. and Kim, B. J. and Keimer, B. and Le Tacon, M.},
  year = 2017,
  month = sep,
  journal = {Physical Review B},
  volume = {96},
  number = {11},
  pages = {115138},
  issn = {2469-9950, 2469-9969},
  doi = {10.1103/PhysRevB.96.115138},
  urldate = {2019-12-12},
  langid = {english}
}

@article{haskelPossibleQuantumParamagnetism2020,
  title = {Possible {{Quantum Paramagnetism}} in {{Compressed Sr}}{\textsubscript{2}}{{IrO}}{\textsubscript{4}}},
  author = {Haskel, D. and Fabbris, G. and Kim, J. H. and Veiga, L. S. I. and Mardegan, J. R. L. and Escanhoela, C. A. and Chikara, S. and Struzhkin, V. and Senthil, T. and Kim, B. J. and Cao, G. and Kim, J.-W.},
  year = 2020,
  month = feb,
  journal = {Physical Review Letters},
  volume = {124},
  number = {6},
  pages = {067201},
  publisher = {American Physical Society},
  doi = {10.1103/PhysRevLett.124.067201},
  urldate = {2021-09-29}
}

@article{jackeliMottInsulatorsStrong2009,
  title = {Mott {{Insulators}} in the {{Strong Spin-Orbit Coupling Limit}}: {{From Heisenberg}} to a {{Quantum Compass}} and {{Kitaev Models}}},
  shorttitle = {Mott {{Insulators}} in the {{Strong Spin-Orbit Coupling Limit}}},
  author = {Jackeli, G. and Khaliullin, G.},
  year = 2009,
  month = jan,
  journal = {Physical Review Letters},
  volume = {102},
  number = {1},
  pages = {017205},
  doi = {10.1103/PhysRevLett.102.017205},
  urldate = {2019-12-10}
}

@article{kimNovel$J_mathrmeff12$2008,
  title = {Novel {{J}}{\textsubscript{eff}}=1/2 {{Mott State Induced}} by {{Relativistic Spin-Orbit Coupling}} in {{SrIr}}{\textsubscript{2}}{{O}}{\textsubscript{4}}},
  author = {Kim, B. J. and Jin, Hosub and Moon, S. J. and Kim, J.-Y. and Park, B.-G. and Leem, C. S. and Yu, Jaejun and Noh, T. W. and Kim, C. and Oh, S.-J. and Park, J.-H. and Durairaj, V. and Cao, G. and Rotenberg, E.},
  year = 2008,
  month = aug,
  journal = {Physical Review Letters},
  volume = {101},
  number = {7},
  pages = {076402},
  publisher = {American Physical Society},
  doi = {10.1103/PhysRevLett.101.076402},
  urldate = {2026-06-22}
}

@article{kimPhaseSensitiveObservationSpinOrbital2009,
  title = {Phase-{{Sensitive Observation}} of a {{Spin-Orbital Mott State}} in {{Sr}}{\textsubscript{2}}{{IrO}}{\textsubscript{4}}},
  author = {Kim, B. J. and Ohsumi, H. and Komesu, T. and Sakai, S. and Morita, T. and Takagi, H. and Arima, T.},
  year = 2009,
  month = mar,
  journal = {Science},
  volume = {323},
  number = {5919},
  pages = {1329--1332},
  publisher = {American Association for the Advancement of Science},
  doi = {10.1126/science.1167106},
  urldate = {2021-10-05}
}

@article{kimQuantumSpinNematic2024,
  title = {Quantum Spin Nematic Phase in a Square-Lattice Iridate},
  author = {Kim, Hoon and Kim, Jin-Kwang and Kwon, Junyoung and Kim, Jimin and Kim, Hyun-Woo J. and Ha, Seunghyeok and Kim, Kwangrae and Lee, Wonjun and Kim, Jonghwan and Cho, Gil Young and Heo, Hyeokjun and Jang, Joonho and Sahle, C. J. and Longo, A. and Strempfer, J. and Fabbris, G. and Choi, Y. and Haskel, D. and Kim, Jungho and Kim, J.-W. and Kim, B. J.},
  year = 2024,
  month = jan,
  journal = {Nature},
  volume = {625},
  number = {7994},
  pages = {264--269},
  publisher = {Nature Publishing Group},
  issn = {1476-4687},
  doi = {10.1038/s41586-023-06829-4},
  urldate = {2024-01-11},
  copyright = {2023 The Author(s), under exclusive licence to Springer Nature Limited},
  langid = {english}
}

@article{kimSingleCrystalGrowth2022,
  title = {Single Crystal Growth of Iridates without Platinum Impurities},
  author = {Kim, Jimin and Kim, Hoon and Kim, Hyun-Woo J. and Park, Sunwook and Kim, Jin-Kwang and Kwon, Junyoung and Kim, Jungho and Seo, Hyeong Woo and Kim, Jun Sung and Kim, B. J.},
  year = 2022,
  month = oct,
  journal = {Physical Review Materials},
  volume = {6},
  number = {10},
  pages = {103401},
  publisher = {American Physical Society},
  doi = {10.1103/PhysRevMaterials.6.103401},
  urldate = {2026-06-03}
}

@article{klimovskikhTunable3D2D2020,
  title = {Tunable {{3D}}/{{2D}} Magnetism in the ({{MnBi}}{\textsubscript{2}}{{Te}}{\textsubscript{4}})({{Bi}}{\textsubscript{2}}{{Te}}{\textsubscript{3}}){\textsubscript{m}} Topological Insulators Family},
  author = {Klimovskikh, Ilya I. and Otrokov, Mikhail M. and Estyunin, Dmitry and Eremeev, Sergey V. and Filnov, Sergey O. and Koroleva, Alexandra and Shevchenko, Eugene and Voroshnin, Vladimir and Rybkin, Artem G. and Rusinov, Igor P. and {Blanco-Rey}, Maria and Hoffmann, Martin and Aliev, Ziya S. and Babanly, Mahammad B. and Amiraslanov, Imamaddin R. and Abdullayev, Nadir A. and Zverev, Vladimir N. and Kimura, Akio and Tereshchenko, Oleg E. and Kokh, Konstantin A. and Petaccia, Luca and Di Santo, Giovanni and Ernst, Arthur and Echenique, Pedro M. and Mamedov, Nazim T. and Shikin, Alexander M. and Chulkov, Eugene V.},
  year = 2020,
  month = aug,
  journal = {npj Quantum Materials},
  volume = {5},
  number = {1},
  pages = {54},
  publisher = {Nature Publishing Group},
  issn = {2397-4648},
  doi = {10.1038/s41535-020-00255-9},
  urldate = {2026-06-03},
  copyright = {2020 The Author(s)},
  langid = {english}
}

@article{laughlinQuantumCriticalityConundrum2001,
  title = {The Quantum Criticality Conundrum},
  author = {Laughlin, R. B. and Lonzarich, G. G. and Monthoux, P. and Pines, David},
  year = 2001,
  month = jun,
  journal = {Advances in Physics},
  volume = {50},
  number = {4},
  pages = {361--365},
  publisher = {Taylor \& Francis},
  issn = {0001-8732},
  doi = {10.1080/00018730110098534},
  urldate = {2026-06-03}
}

@article{liMagneticOrderDisorder2021,
  title = {Magnetic Order, Disorder, and Excitations under Pressure in the {{Mott}} Insulator {{Sr}}{\textsubscript{2}}{{IrO}}{\textsubscript{4}}},
  author = {Li, Xiang and Cooper, S. E. and Krishnadas, A. and {de la Torre}, A. and Perry, R. S. and Baumberger, F. and Silevitch, D. M. and Hsieh, D. and Rosenbaum, T. F. and Feng, Yejun},
  year = 2021,
  month = nov,
  journal = {Physical Review B},
  volume = {104},
  number = {20},
  pages = {L201111},
  publisher = {American Physical Society},
  doi = {10.1103/PhysRevB.104.L201111},
  urldate = {2021-11-20}
}

@article{liOpticalRamanMeasurements2020,
  title = {Optical {{Raman}} Measurements of Low Frequency Magnons under High Pressure},
  author = {Li, Xiang and Cooper, S. E. and Krishnadas, A. and Silevitch, D. M. and Rosenbaum, T. F. and Feng, Yejun},
  year = 2020,
  month = nov,
  journal = {Review of Scientific Instruments},
  volume = {91},
  number = {11},
  pages = {113902},
  publisher = {American Institute of Physics},
  issn = {0034-6748},
  doi = {10.1063/5.0026311},
  urldate = {2020-11-04}
}

@article{liPressurecontrolledInterlayerMagnetism2019,
  title = {Pressure-Controlled Interlayer Magnetism in Atomically Thin {{CrI}}{\textsubscript{3}}},
  author = {Li, Tingxin and Jiang, Shengwei and Sivadas, Nikhil and Wang, Zefang and Xu, Yang and Weber, Daniel and Goldberger, Joshua E. and Watanabe, Kenji and Taniguchi, Takashi and Fennie, Craig J. and Fai Mak, Kin and Shan, Jie},
  year = 2019,
  month = dec,
  journal = {Nature Materials},
  volume = {18},
  number = {12},
  pages = {1303--1308},
  publisher = {Nature Publishing Group},
  issn = {1476-4660},
  doi = {10.1038/s41563-019-0506-1},
  urldate = {2026-06-03},
  copyright = {2019 The Author(s), under exclusive licence to Springer Nature Limited},
  langid = {english}
}

@article{padmanabhanInterlayerMagnetophononicCoupling2022,
  title = {Interlayer Magnetophononic Coupling in {{MnBi}}{\textsubscript{2}}{{Te}}{\textsubscript{4}}},
  author = {Padmanabhan, Hari and Poore, Maxwell and Kim, Peter K. and Koocher, Nathan Z. and Stoica, Vladimir A. and Puggioni, Danilo and (Hugo) Wang, Huaiyu and Shen, Xiaozhe and Reid, Alexander H. and Gu, Mingqiang and Wetherington, Maxwell and Lee, Seng Huat and Schaller, Richard D. and Mao, Zhiqiang and Lindenberg, Aaron M. and Wang, Xijie and Rondinelli, James M. and Averitt, Richard D. and Gopalan, Venkatraman},
  year = 2022,
  month = apr,
  journal = {Nature Communications},
  volume = {13},
  number = {1},
  pages = {1929},
  publisher = {Nature Publishing Group},
  issn = {2041-1723},
  doi = {10.1038/s41467-022-29545-5},
  urldate = {2026-06-03},
  copyright = {2022 The Author(s)},
  langid = {english}
}

@article{porrasPseudospinlatticeCouplingSpinorbit2019,
  title = {Pseudospin-Lattice Coupling in the Spin-Orbit {{Mott}} Insulator {{Sr}}{\textsubscript{2}}{{IrO}}{\textsubscript{4}}},
  author = {Porras, J. and Bertinshaw, J. and Liu, H. and Khaliullin, G. and Sung, N. H. and Kim, J.-W. and Francoual, S. and Steffens, P. and Deng, G. and Sala, M. Moretti and Efimenko, A. and Said, A. and Casa, D. and Huang, X. and Gog, T. and Kim, J. and Keimer, B. and Kim, B. J.},
  year = 2019,
  month = feb,
  journal = {Physical Review B},
  volume = {99},
  number = {8},
  pages = {085125},
  publisher = {American Physical Society},
  doi = {10.1103/PhysRevB.99.085125},
  urldate = {2021-06-03}
}

@article{savaryDisorderInducedQuantumSpin2017,
  title = {Disorder-{{Induced Quantum Spin Liquid}} in {{Spin Ice Pyrochlores}}},
  author = {Savary, Lucile and Balents, Leon},
  year = 2017,
  month = feb,
  journal = {Physical Review Letters},
  volume = {118},
  number = {8},
  pages = {087203},
  publisher = {American Physical Society},
  doi = {10.1103/PhysRevLett.118.087203},
  urldate = {2026-06-03}
}

@article{songSwitching2DMagnetic2019,
  title = {Switching {{2D}} Magnetic States via Pressure Tuning of Layer Stacking},
  author = {Song, Tiancheng and Fei, Zaiyao and Yankowitz, Matthew and Lin, Zhong and Jiang, Qianni and Hwangbo, Kyle and Zhang, Qi and Sun, Bosong and Taniguchi, Takashi and Watanabe, Kenji and McGuire, Michael A. and Graf, David and Cao, Ting and Chu, Jiun-Haw and Cobden, David H. and Dean, Cory R. and Xiao, Di and Xu, Xiaodong},
  year = 2019,
  month = dec,
  journal = {Nature Materials},
  volume = {18},
  number = {12},
  pages = {1298--1302},
  publisher = {Nature Publishing Group},
  issn = {1476-4660},
  doi = {10.1038/s41563-019-0505-2},
  urldate = {2026-06-03},
  copyright = {2019 The Author(s), under exclusive licence to Springer Nature Limited},
  langid = {english}
}

@article{stahlPressuretuningARuCl3Quantum2024,
  title = {Pressure-Tuning of {$\alpha$}-{{RuCl}}{\textsubscript{3}} towards a Quantum Spin Liquid},
  author = {Stahl, Q. and Ritschel, T. and Garbarino, G. and Cova, F. and Isaeva, A. and Doert, T. and Geck, J.},
  year = 2024,
  month = sep,
  journal = {Nature Communications},
  volume = {15},
  number = {1},
  pages = {8142},
  publisher = {Nature Publishing Group},
  issn = {2041-1723},
  doi = {10.1038/s41467-024-52169-w},
  urldate = {2026-06-22},
  copyright = {2024 The Author(s)},
  langid = {english}
}

@article{takayamaModelAnalysisMagnetic2016,
  title = {Model Analysis of Magnetic Susceptibility of {{Sr}}{\textsubscript{2}}{{IrO}}{\textsubscript{4}}: {{A}} Two-Dimensional {{J}}{\textsubscript{eff}} = 1/2 {{Heisenberg}} System with Competing Interlayer Couplings},
  shorttitle = {Model Analysis of Magnetic Susceptibility of {{Sr}} 2 {{IrO}} 4},
  author = {Takayama, Tomohiro and Matsumoto, Akiyo and Jackeli, George and Takagi, Hidenori},
  year = 2016,
  month = dec,
  journal = {Physical Review B},
  volume = {94},
  number = {22},
  pages = {224420},
  issn = {2469-9950, 2469-9969},
  doi = {10.1103/PhysRevB.94.224420},
  urldate = {2024-02-09},
  langid = {english}
}

@article{vlaskovaMagneticPropertiesCrystal2020,
  title = {Magnetic Properties and Crystal Field Splitting of the Rare-Earth Pyrochlore {{Er}}{\textsubscript{2}}{{Ir}}{\textsubscript{2}}{{O}}{\textsubscript{7}}},
  author = {Vl{\'a}{\v s}kov{\'a}, Kristina and Proschek, Petr and Divi{\v s}, Martin and Le, Duc and Colman, Ross Harvey and Klicpera, Milan},
  year = 2020,
  month = aug,
  journal = {Physical Review B},
  volume = {102},
  number = {5},
  pages = {054428},
  publisher = {American Physical Society},
  doi = {10.1103/PhysRevB.102.054428},
  urldate = {2021-01-27}
}

@article{yeMagneticCrystalStructures2013,
  title = {Magnetic and Crystal Structures of {{Sr}}{\textsubscript{2}}{{IrO}}{\textsubscript{4}}: {{A}} Neutron Diffraction Study},
  shorttitle = {Magnetic and Crystal Structures of \$\textbraceleft\textbackslash mathrm\textbraceleft{{Sr}}\textbraceright\textbraceright\_\textbraceleft 2\textbraceright\textbraceleft\textbackslash mathrm\textbraceleft{{IrO}}\textbraceright\textbraceright\_\textbraceleft 4\textbraceright\$},
  author = {Ye, Feng and Chi, Songxue and Chakoumakos, Bryan C. and {Fernandez-Baca}, Jaime A. and Qi, Tongfei and Cao, G.},
  year = 2013,
  month = apr,
  journal = {Physical Review B},
  volume = {87},
  number = {14},
  pages = {140406R},
  publisher = {American Physical Society},
  doi = {10.1103/PhysRevB.87.140406},
  urldate = {2021-10-05}
}

@article{yeStructureSymmetryDetermination2015,
  title = {Structure Symmetry Determination and Magnetic Evolution in {{Sr}}{\textsubscript{2}}{{Ir}}{\textsubscript{x}}{{Rh}}{\textsubscript{1-x}}{{O}}{\textsubscript{4}}},
  author = {Ye, Feng and Wang, Xiaoping and Hoffmann, Christina and Wang, Jinchen and Chi, Songxue and Matsuda, Masaaki and Chakoumakos, Bryan C. and {Fernandez-Baca}, Jaime A. and Cao, G.},
  year = 2015,
  month = nov,
  journal = {Physical Review B},
  volume = {92},
  number = {20},
  pages = {201112},
  publisher = {American Physical Society},
  doi = {10.1103/PhysRevB.92.201112},
  urldate = {2026-02-19}
}

@article{Zocco:2014bd,
  title = {Persistent Non-Metallic Behavior in {{Sr}}{\textsubscript{2}}{{IrO}}{\textsubscript{4}} and {{Sr}}{\textsubscript{3}}{{Ir}}{\textsubscript{2}}{{O}}{\textsubscript{7}} at High Pressures},
  author = {Zocco, D A and Hamlin, J J and White, B D and Kim, B J and Jeffries, J R and Weir, S T and Vohra, Y K and Allen, J W and Maple, M B},
  year = 2014,
  month = jun,
  journal = {Journal of Physics: Condensed Matter},
  volume = {26},
  number = {25},
  pages = {255603--6},
  doi = {10.1088/0953-8984/26/25/255603},
  uri = {papers3://publication/doi/10.1088/0953-8984/26/25/255603}
}

@misc{YePrivate2026,
  title = {Private Communication},
  author = {Ye, Feng},
  year = 2026
}

@article{parisStrainEngineeringCharge2020,
    chapter = {Physical Sciences},
    title = {Strain engineering of the charge and spin-orbital interactions in {Sr}$_{\textrm{2}}${IrO}$_{\textrm{4}}$},
    copyright = {Copyright © 2020 the Author(s). Published by PNAS.. https://creativecommons.org/licenses/by-nc-nd/4.0/This open access article is distributed under Creative Commons Attribution-NonCommercial-NoDerivatives License 4.0 (CC BY-NC-ND).},
    issn = {0027-8424, 1091-6490},
    url = {https://www.pnas.org/content/early/2020/09/16/2012043117},
    doi = {10.1073/pnas.2012043117},
    urldate = {2020-09-21},
    journal = {Proceedings of the National Academy of Sciences},
    publisher = {National Academy of Sciences},
    author = {Paris, Eugenio and Tseng, Yi and Pärschke, Ekaterina M. and Zhang, Wenliang and Upton, Mary H. and Efimenko, Anna and Rolfs, Katharina and McNally, Daniel E. and Maurel, Laura and Naamneh, Muntaser and Caputo, Marco and Strocov, Vladimir N. and Wang, Zhiming and Casa, Diego and Schneider, Christof W. and Pomjakushina, Ekaterina and Wohlfeld, Krzysztof and Radovic, Milan and Schmitt, Thorsten},
    month = sep,
    year = {2020},
}

@article{boseggiaLockingIridiumMagnetic2013,
  title = {Locking of Iridium Magnetic Moments to the Correlated Rotation of Oxygen Octahedra in {Sr}$_{\textrm{2}}${IrO}$_{\textrm{4}}$ Revealed by X-Ray Resonant Scattering},
  author = {Boseggia, S and Walker, H C and Vale, J and Springell, R and Feng, Z and Perry, R S and Moretti Sala, M and R{\o}nnow, H M and Collins, S P and McMorrow, D F},
  year = 2013,
  month = sep,
  journal = {Journal of Physics: Condensed Matter},
  volume = {25},
  number = {42},
  pages = {422202},
  publisher = {IOP Publishing},
  issn = {0953-8984},
  doi = {10.1088/0953-8984/25/42/422202},
  urldate = {2026-07-10},
  langid = {english}
}

@article{pinciniAnisotropicExchangeSpinwave2017,
  title = {Anisotropic Exchange and Spin-Wave Damping in Pure and Electron-Doped {Sr}$_{\textrm{2}}${IrO}$_{\textrm{4}}$},
  author = {Pincini, D. and Vale, J. G. and Donnerer, C. and {de la Torre}, A. and Hunter, E. C. and Perry, R. and Moretti Sala, M. and Baumberger, F. and McMorrow, D. F.},
  year = 2017,
  month = aug,
  journal = {Physical Review B},
  volume = {96},
  number = {7},
  pages = {075162},
  publisher = {American Physical Society},
  doi = {10.1103/PhysRevB.96.075162},
  urldate = {2026-07-10}
}

@article{Cata4444Xu_2020,
   title={High-throughput calculations of magnetic topological materials},
   volume={586},
   ISSN={1476-4687},
   url={http://dx.doi.org/10.1038/s41586-020-2837-0},
   DOI={10.1038/s41586-020-2837-0},
   number={7831},
   journal={Nature},
   publisher={Springer Science and Business Media LLC},
   author={Xu, Yuanfeng and Elcoro, Luis and Song, Zhi-Da and Wieder, Benjamin J. and Vergniory, M. G. and Regnault, Nicolas and Chen, Yulin and Felser, Claudia and Bernevig, B. Andrei},
   year={2020},
   month=10, pages={702–707} }

@article{karaki2024highthroughputsearchtopologicalmagnon,
   title={High-throughput discovery of perturbation-induced topological magnons},
   volume={11},
   ISSN={2057-3960},
   url={http://dx.doi.org/10.1038/s41524-025-01706-2},
   DOI={10.1038/s41524-025-01706-2},
   number={1},
   journal={npj Computational Materials},
   publisher={Springer Science and Business Media LLC},
   author={Karaki, Mohammed J. and Fahmy, Ahmed E. and Williams, Archibald J. and Haravifard, Sara and Goldberger, Joshua E. and Lu, Yuan-Ming},
   year={2025},
   month=jul }

@article{MTQCElcoro_2021,
   title={Magnetic topological quantum chemistry},
   volume={12},
   ISSN={2041-1723},
   url={http://dx.doi.org/10.1038/s41467-021-26241-8},
   DOI={10.1038/s41467-021-26241-8},
   number={1},
   journal={Nature Communications},
   publisher={Springer Science and Business Media LLC},
   author={Elcoro, Luis and Wieder, Benjamin J. and Song, Zhida and Xu, Yuanfeng and Bradlyn, Barry and Bernevig, B. Andrei},
   year={2021},
   month=10 }

@article{GretarssonPhysRevLett.117.107001,
  title = {Persistent Paramagnons Deep in the Metallic Phase of {Sr}$_{2-x}${La}$_x${IrO}$_{\textrm{4}}$},
  author = {Gretarsson, H. and Sung, N. H. and Porras, J. and Bertinshaw, J. and Dietl, C. and Bruin, Jan A. N. and Bangura, A. F. and Kim, Y. K. and Dinnebier, R. and Kim, Jungho and Al-Zein, A. and Moretti Sala, M. and Krisch, M. and Le Tacon, M. and Keimer, B. and Kim, B. J.},
  journal = {Phys. Rev. Lett.},
  volume = {117},
  issue = {10},
  pages = {107001},
  numpages = {6},
  year = {2016},
  month = {Sep},
  publisher = {American Physical Society},
  doi = {10.1103/PhysRevLett.117.107001},
  url = {https://link.aps.org/doi/10.1103/PhysRevLett.117.107001}
}

@article{torchinskyStructuralDistortionInducedMagnetoelastic2015,
  title = {Structural {{Distortion-Induced Magnetoelastic Locking}} in {{Sr}}{\textsubscript{2}}{{IrO}}{\textsubscript{4}} {{Revealed}} through {{Nonlinear Optical Harmonic Generation}}},
  author = {Torchinsky, D. H. and Chu, H. and Zhao, L. and Perkins, N. B. and Sizyuk, Y. and Qi, T. and Cao, G. and Hsieh, D.},
  year = 2015,
  month = mar,
  journal = {Physical Review Letters},
  volume = {114},
  number = {9},
  pages = {096404},
  publisher = {American Physical Society},
  doi = {10.1103/PhysRevLett.114.096404},
  urldate = {2026-08-14}
}

\end{document}